\documentclass[twocolumn]{htl-author}
\usepackage{color}
\usepackage{ulem}

\DOI{2012.0357}

\begin{document}

\title{Realistic Endoscopic Image Generation Method \\Using Virtual-to-real Image-domain Translation}

\author{Masahiro Oda$^{1}$, Kiyohito Tanaka$^{2}$, Hirotsugu Takabatake$^{3}$, Masaki Mori$^{4}$,\\ Hiroshi Natori$^{5}$, and Kensaku Mori$^{1,6}$}

\address{$^{1}$Graduate School of Informatics, Nagoya University, Furo-cho, Chikusa-ku, Nagoya, Aichi, 464-8601, Japan\\
E-mail: moda@mori.m.is.nagoya-u.ac.jp\\
$^{2}$Department of Gastroenterology, Kyoto Second Red Cross Hospital, \\355-5, Haruobi-cho, Kamigyo-ku, Kyoto, Kyoto, 602-8026, Japan\\
$^{3}$Department of Respiratory Medicine, Sapporo-Minami-Sanjo Hospital, \\6, Nishi, Minami-3-jo, Chuo-ku, Sapporo, Hokkaido, 060-0063, Japan\\
$^{4}$Department of Respiratory Medicine, Sapporo-Kosei General Hospital, \\8-5, Higashi, Kita-3-jo, Chuo-ku, Sapporo, Hokkaido, 060-0033, Japan\\
$^{5}$Department of Respiratory Medicine, Keiwakai Nishioka Hospital, \\1-52, Nishioka-4-jo 4-chome, Toyohira-ku, Sapporo, Hokkaido, 062-0034, Japan\\
$^{6}$Research Center for Medical Bigdata, National Institute of Informatics, \\2-1-2 Hitotsubashi, Chiyoda-ku, Tokyo, 101-8430, Japan}

%\historydate{Published in Healthcare Technology Letters; Received on xxx; Revised on xxx}

%200 words
\abstract{This paper proposes a realistic image generation method for visualization in endoscopic simulation systems.
Endoscopic diagnosis and treatment are performed in many hospitals.
To reduce complications related to endoscope insertions, endoscopic simulation systems are used for training or rehearsal of endoscope insertions.
However, current simulation systems generate non-realistic virtual endoscopic images.
To improve the value of the simulation systems, improvement of reality of their generated images is necessary.
We propose a realistic image generation method for endoscopic simulation systems.
Virtual endoscopic images are generated by using a volume rendering method from a CT volume of a patient.
We improve the reality of the virtual endoscopic images using a virtual-to-real image-domain translation technique.
The image-domain translator is implemented as a fully convolutional network (FCN).
We train the FCN by minimizing a cycle consistency loss function.
The FCN is trained using unpaired virtual and real endoscopic images.
To obtain high quality image-domain translation results, we perform an image cleansing to the real endoscopic image set.
We tested to use the shallow U-Net, U-Net, deep U-Net, and U-Net having residual units as the image-domain translator.
The deep U-Net and U-Net having residual units generated quite realistic images.
}

\maketitle

\section{Introduction}\label{sec:intro}

Endoscopic diagnosis and treatment are commonly performed in hospitals.
An endoscope is inserted into a patient's body through an incised part or a natural orifice to capture interior views of the body.
Endoscopes such as a colonoscope, gastroendoscope, and bronchoscope are used to diagnose digestive organs, trachea, and bronchi.
Laparoscope and thoracoscope are used in surgeries.
In endoscopic diagnosis and treatment, a surgeon or a physician needs to understand condition of target organs or lesions from endoscopic images.
Because of the limitation of the field of view and difficulty of understanding 3D structure, endoscopic procedures are difficult to perform.
Complications occur in endoscopic diagnosis and treatment due to their difficulty.
In colon diagnosis and treatment, some inexperienced physicians cause colon perforation using colonoscope.
In laparoscopic surgeries, some inexperienced surgeons injures organs by the laparoscope or other surgical tools outside of the viewing fields of the laparoscope.

Complications caused related to insufficient recognitions of organ structures by physicians or surgeons during endoscopic procedures need to be reduced.
To reduce such complications, trainings of endoscopic procedures using simulation systems are performed \cite{Miki16,khan18,Triantafyllou14,Harpham15}.
Endoscopic simulation systems are useful because physicians or surgeons can improve their manipulation ability without any trouble with patients.
Such simulation systems are effective for less-experienced physicians or surgeons.
When pre-operative 3D images of a patient is available, a patient-specific simulation or rehearsal can be performed using an endoscopic simulation system.
Such patient-specific simulation systems are effective not only for less-experienced but also experienced physicians or surgeons.

One problem of endoscopic simulation systems is the lack of reality on their generated virtual endoscopic images.
Commonly, textures of the organ surfaces and light reflection on the virtual endoscopic images are different from the real images.
The simulation systems commonly generate virtual images from surface or volumetric model of organs.
The surface or volumetric model represents anatomical structures coarsely compared to the real organ shapes.
Detailed anatomical structures such as tiny blood vessels on organ surfaces are not included in the models.
Such tiny blood vessels affect color and texture of the organ surfaces.
Color and textures of the organ surfaces in real images need to be considered in image generations.
To improve reality of virtual images, the texture mapping technique is used in simulation systems.
However, preparation of good texture images for the texture mapping is difficult.
Furthermore, the light reflection model employed by the simulation systems cannot simulate the real light reflection due to lack of the light reflection/absorption property information of the organ surfaces.
The appearance of light reflection/absorption on the organ surfaces is not realistic on the generated images.
In summary, (1) regardless of real textures of the organ surfaces and (2) lack of the light reflection/absorption property information of the organ surfaces reduce reality of the generated virtual endoscopic images.

To improve the value of endoscopic simulation systems, improvement of reality of their generated virtual endoscopic images is necessary.
Simulation systems having realistic image renderer greatly contribute skill improvement of surgeons or physicians in trainings or rehearsals of endoscope manipulation.

Recently, generative adversarial networks (GANs) are utilized for realistic image generation.
Shrivastava et al. \cite{Shrivastava16} improved reality of simulated images using a framework of GAN to generate realistic images having annotation information.
They applied their method to eye images and distance images of hand pose.
Engelhardt et al. \cite{Engelhardt18} proposed a method to improve reality of endoscopic images taken from surgical phantoms.
They used the tempCycleGAN to convert endoscopic images.
Gil et al. \cite{Gil19} and Esteban-Lansaque et al. \cite{Esteban19} improved reality of virtual bronchoscopic images using GANs.
Their ideas are mainly based on CycleGAN \cite{cyclegan}.
CycleGAN enables establishment of image translator from unpaired data.
CycleGAN-based methods were proposed not only for realistic image generation but also segmentation \cite{Huo18} and image modality conversion \cite{Wolterink17,Hiasa18}.
The previous method \cite{Engelhardt18} introduced temporal information to CycleGAN to achieve temporally consistent results.
Also, some methods \cite{Huo18,Wolterink17,Hiasa18} introduced original loss functions to impose constraint conditions to CycleGAN.
CycleGAN uses an image-domain translation network to perform image conversion.
The structure of the network is strongly related to its ability of learning representations from images.
Quality of image-domain translation is significantly affected by the network.
However, the relationship between image-domain translation quality and network structure has not been investigated.

We propose a realistic image generation method for endoscopic simulation systems.
We generate virtual endoscopic images from a CT volume of a patient.
We translate the virtual endoscopic images into the real endoscopic image domain using a virtual-to-real image-domain translation technique.
The image-domain translator is build based on a large number of virtual and real endoscopic images as a fully convolutional network (FCN).
Our method can be trained using unpaired image data.
To obtain high quality image-domain translator, we perform an image cleansing to the real endoscopic image set.
The quality of translated images depends on the FCN network structure of the image-domain translator.
We employed network structures including the shallow U-Net, U-Net, deep U-Net, and U-Net having residual units as the image-domain translator.
Relationships between the network structures and translated image quality are investigated in our experiments.
Our method can automatically generate realistic textures on the organ surfaces in the virtual endoscopic images.

The contributions of this paper are (1) proposal of realistic virtual endoscopic image generation method and (2) investigation of relationships between image-domain translation network structures and quality of translated images.

\section{Realistic Endoscopic Image Generation Method}\label{sec:method}

\subsection{Overview}

The proposed method converts virtual endoscopic images into realistic endoscopic images.
We use sets of the virtual and real endoscopic images.
These images are obtained from different patients (unpaired data).
In the real endoscopic image set, some images that are not suitable to establish translation from virtual to real domain.
Samples of such images and the reason why they are not suitable are given in \ref{sssec:realimage}.
Therefore, we apply "a data cleansing" process to improve the quality of image-domain translation.
We make an image-domain translation network based on the CycleGAN \cite{cyclegan} technique.
The image-domain translation network is trained to minimize a cycle consistency loss.
We used four FCNs as the image-domain translator to investigate relationships between network structures and image-domain translation quality.

\subsection{Data preparation and cleansing}

\subsubsection{Virtual endoscopic images}

\begin{figure}[!b]
\centering{\includegraphics[width=0.48\textwidth]{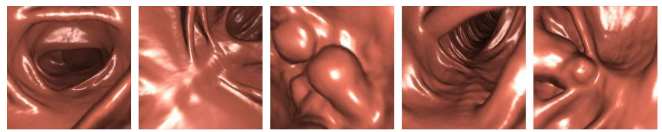}}
\caption{Samples of virtual endoscopic images taken in colon. \label{fig:virtualimage}}
\end{figure}

We generate virtual endoscopic images from CT volumes using a volume rendering method \cite{Mori03}.
These images were taken during manual fly-through in hollow organs.
Samples of images taken in the colon are shown in Fig. \ref{fig:virtualimage}.
The set of virtual endoscopic images are denoted as $V$.

\subsubsection{Real endoscopic images} \label{sssec:realimage}

\begin{figure}[!b]
\centering{\includegraphics[width=0.48\textwidth]{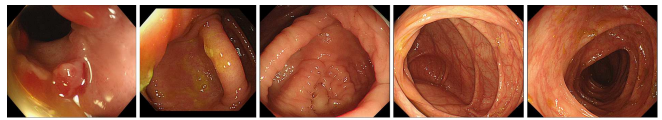}}
\caption{Samples of real endoscopic images taken in colon. \label{fig:realimage}}
\end{figure}

\begin{figure}[!b]
\centering{\includegraphics[width=0.48\textwidth]{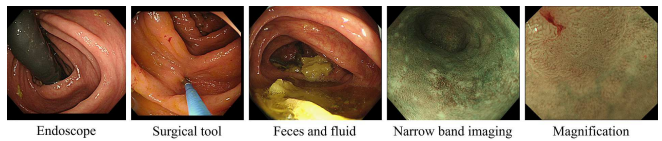}}
\caption{Samples of real endoscopic images that are not used to train image-domain translator. These images contain endoscope, surgical tool, feces, and fluid. Also, images taken in narrow band imaging and magnification modes of endoscopes are not used to train.
\label{fig:realimage_notusedsample}}
\end{figure}

We capture real endoscopic images during endoscope insertions to patients (Fig. \ref{fig:realimage}).
Among these images, we exclude images that are not suitable for the image-domain translation.
This process is called as "data cleansing" process.
This process contributes to improve the quality of image-domain translation.
To obtain better image-domain translation results, both of the images in the source and target domains should capture similar objects, such as the organs.
However, some real endoscopic images contain parts of endoscope, surgical tools, feces, and fluid.
Also, some real endoscopic images are taken in special imaging modes of endoscopes such as the narrow band imaging and magnification modes.
Samples of such images are shown in Fig. \ref{fig:realimage_notusedsample}.
They never appear in images in the virtual domain.
Therefore, we remove these images from the set of real endoscopic images used to train the image-domain translator.
The set of the remaining real endoscopic image are denoted as $R$.

\subsection{Virtual-to-real image-domain translation}

\subsubsection{Image-domain translation method}

We make virtual-to-real image-domain translation by using the sets of images $V$ and $R$.
%Sets of images in virtual and real endoscopic image domains are denoted as $V$ and $R$, respectively.
The virtual-to-real image-domain translator performs mapping $G : V \rightarrow R$.
We achieve the mapping $G$ in adversarial training frameworks.
An adversarial discriminator $D_{R}$, which discriminates between real images $r$ and translated real images $G(v)$.
$v$ is a virtual image in $V$ and $r$ is a real image in $R$.
Based on adversarial loss \cite{goodfellow14}, we define an objective function for mapping $G$ and discriminator $D_{R}$ as
\begin{eqnarray}
L_{gan}(G, D_{R}, V, R) &=& E_{r \in R}\{ {\rm log}D_{R}(r) \} + \\ \nonumber
& & E_{v \in V} \{ {\rm log}(1-D_{R}(G(v))) \}.
\end{eqnarray}
When real images are provided, the function gives likelihood $D_{R}(r) \in [0,1]$.
When virtual images are provided, the images are translated as $G(v)$, and the function gives likelihood $1-D_{R}(G(v)) \in [0,1]$.
The objective function gives higher values if $D_{R}$ discriminates real and translated real images correctly.
Also, the objective function gives lower values if $G$ generates good translated real images that fool the discriminator.
$G$ and $D_{R}$ are trained to achieve $\arg \min_{G} \max_{D_{R}} L_{gan}(G, D_{R}, V, R)$.
Similarly, we define an inverse mapping $F : R \rightarrow V$.
An objective function for mapping $F$ and discriminator $D_{V}$, which discriminates between $v$ and $F(r)$, is also defined as $L_{gan}(F, D_{V}, R, V)$.
However, use of the objective function is not enough to limit range of distributions of images in a target domain.
Appropriate limitation to the distribution of a target domain makes translation results better.
To achieve better mapping $G$, we introduce a cycle consistency loss as proposed in the cycle-consistent adversarial networks \cite{cyclegan}.
The loss implies an image-translation result of an input image by $G$ and $F$ is similar to the input image.
%分かりにくい？
This means $v \rightarrow G(v) \rightarrow F(G(v)) \approx v$.
The cycle consistency loss is defined as
\begin{eqnarray}
L_{cyc}(G, F) &=& E_{v \in V}\{||F(G(v)) - v||_{1}\} + \\ \nonumber
& & E_{r \in R}\{||G(F(r))-r||_{1}\}.
\end{eqnarray}
We define overall objective function as
\begin{eqnarray}
L(G, F, D_{V}, D_{R}) &=& L_{gan}(G, D_{R}, V, R) + \\ \nonumber
& & L_{gan}(F, D_{V}, R, V) + \\ \nonumber
& & \lambda L_{cyc}(G, F),
\end{eqnarray}
where $\lambda$ is the weight of the cycle consistency loss.
We train $G, F, D_{V}, D_{R}$ to achieve $\arg \min_{G,F} \max_{D_{V},D_{R}} L(G, F, D_{V}, D_{R})$.

For the training, virtual endoscopic images $v_{i} (i=1, \ldots, I) \in V$ and real endoscopic images $r_{j} (j=1,\dots, J) \in R$ are used.
$I$ and $J$ are the numbers of virtual and real endoscopic images.
%All of the images are $256 \times 256$ pixels color images.
%Both $G$ and $F$ are implemented as U-nets \cite{unet} having instance normalization \cite{instancenormalization} after each convolution layer.
Both the discriminators $D_{R}$ and $D_{V}$ are implemented as convolutional neural networks.
Both $G$ and $F$ are implemented as FCNs.
Details of the network structures of $G$ and $F$ are described in \ref{ssec:network}.
After the training, we obtain $G$ as the virtual-to-real endoscopic image-domain translator.
%Examples of image-domain translation results by a trained $G$ are shown in Fig. \ref{fig:domaintranslationresult}.

\subsubsection{Network structure of image-domain translator}\label{ssec:network}

The quality of images generated by the image-domain translator is affected by network structures of them.
Therefore, we investigate relationships between the network structures and image-domain translation results.
We use networks including the {\it shallow U-Net}, {\it U-Net}, {\it deep U-Net}, and {\it U-Net having residual units} as $G$ and $F$.
Each structure is explained below.

{\it The U-Net} used here is a modified version of the original U-Net \cite{unet}, which has the instance normalization \cite{instancenormalization} after each convolution layer.

{\it The shallow U-Net} is made by removing layers in the U-Net.
A convolution layer is removed from the encoding and decoding paths of the U-Net, respectively.
%ネットワークの図かく

{\it The deep U-Net} is made by adding layers to the U-Net.
A convolution layer is added to the encoding and decoding paths of the U-Net, respectively.

{\it The U-Net having residual units} is made by replacing convolution layers in the U-Net with residual units \cite{resnet}.

\section{Experiments}\label{sec:exp}

We performed realistic endoscopic image generation using the proposed method.
We used virtual and real colonoscopic images in this experiment.
We evaluated generated images qualitatively.

To train the virtual-to-real image-domain translator, we generated $I=8072$ virtual colonoscopic images from six cases of abdominal CT volumes taken for CT colonography diagnosis (air is insufflated to the colon).
The virtual colonoscopic images were captured during manual fly-through in the colons.
We collected 18775 real colonoscopic images, which were captured during colonoscope insertions to patients.
Two colonoscopists operated a colonoscope.
The real colonoscopic images were captured at random intervals during colonoscope insertions (the images are not consecutive video frames).
The data cleansing process was applied to the images.
5369 images were removed from the images in the process.
The remaining 13406 images were obtained as the data cleansed images.
These images were used to train the image-domain translator.
The CT volumes and the real colonoscopic images were taken from different patients.
The networks were trained in 100 epochs.
100 epochs were enough for training the networks because the large numbers of images were provided in one epoch in trainings.
The mini batch size was 20.
The size of all images used in the proposed method was $256 \times 256$ pixels.

We performed training of the networks on a NVIDIA Quadro P6000 because major amounts of GPU memory were necessary to load images in mini batches. After the training, we performed image-domain translations using the trained networks on a NVIDIA TITAN V.

To confirm how the data cleansing process contributes to improve the quality of the image-domain translation results, we trained the image-domain translator using following two datasets.
\begin{description}
\item[Cleansing:] 13406 real colonoscopic images (data cleansing was applied) and 8072 virtual colonoscopic images.
\item[No cleansing:] 18775 real colonoscopic images (data cleansing was not applied) and 8072 virtual colonoscopic images.
\end{description}
We compared the two image-domain translators trained using the two datasets.

\section{Results}\label{sec:result}

\begin{figure*}[!b]
\centering{\includegraphics[width=0.98\textwidth]{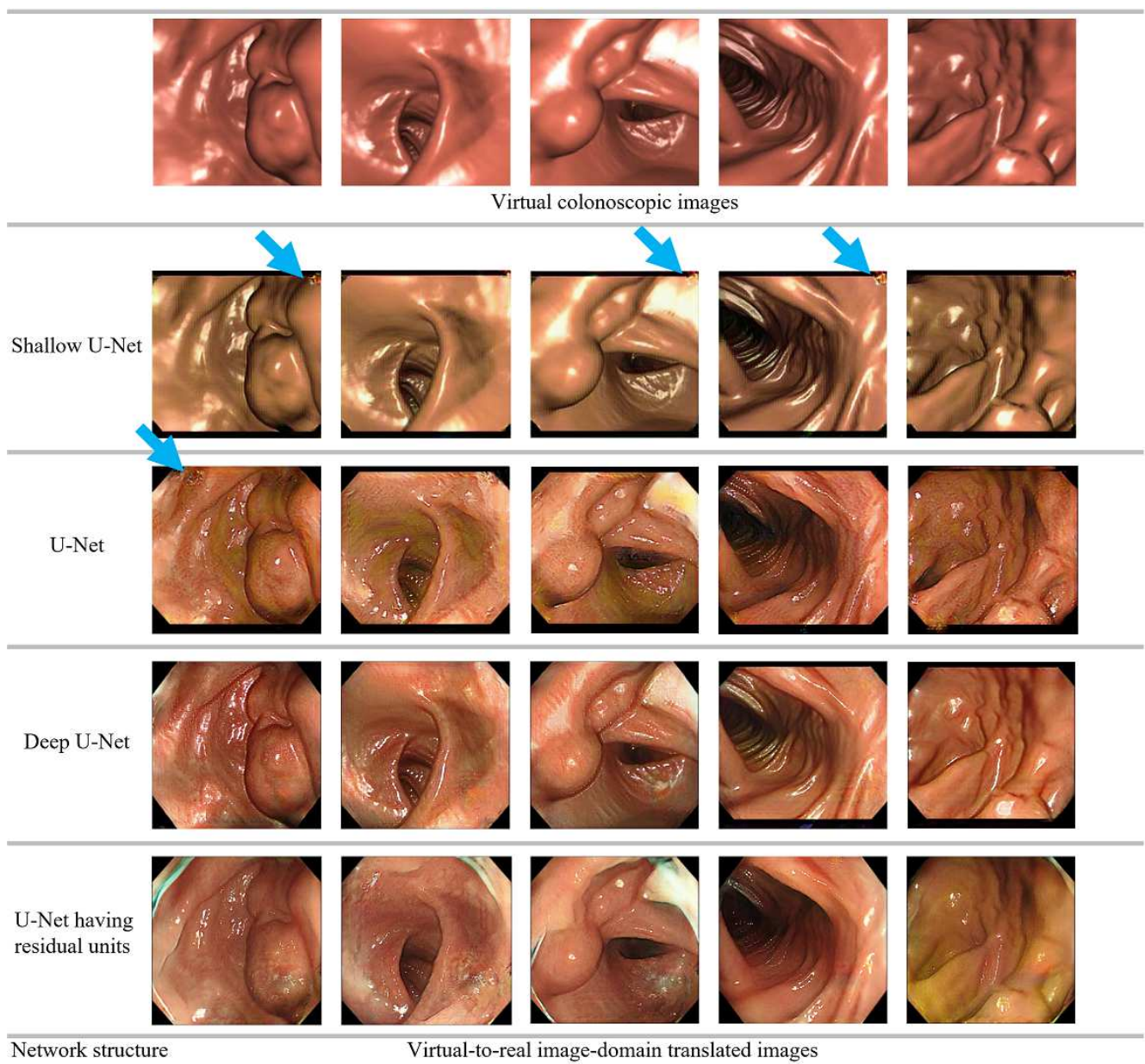}}
\caption{Top row shows virtual colonoscopic images. These images were given to image-domain translator trained using "Cleansing" dataset. Remaining rows show image-domain translation results obtained using shallow U-Net, U-Net, deep U-Net, and U-Net having residual units. Unnatural parts in translated images are indicated by arrows. \label{fig:result}}
\end{figure*}

\begin{figure*}[tb]
\begin{center}
\begin{tabular}{c}
\includegraphics[width=0.98\textwidth]{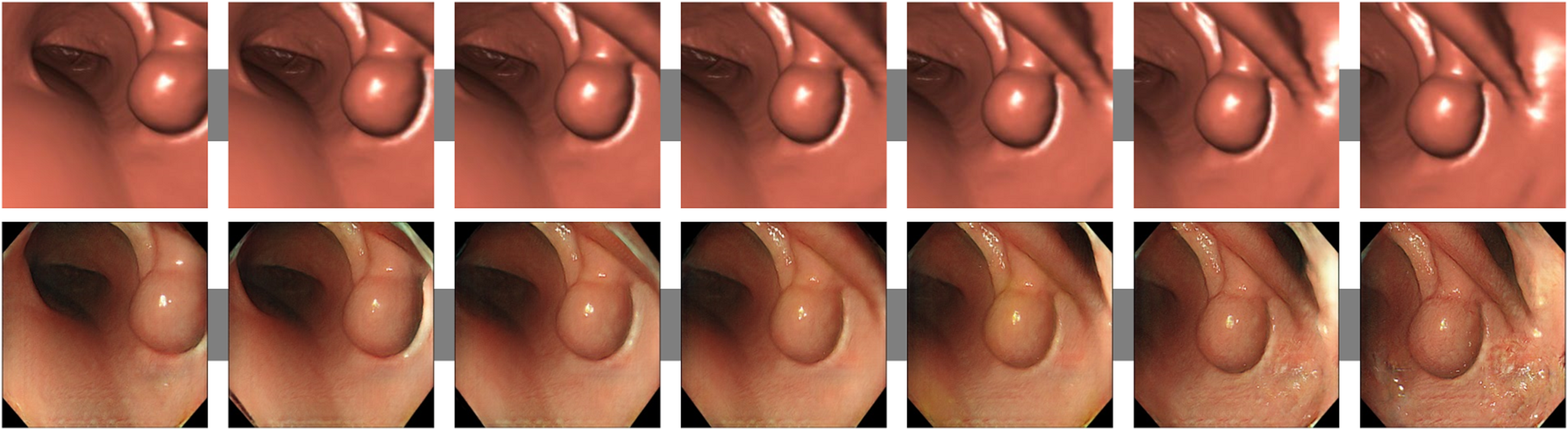} \\
(a) \\
\includegraphics[width=0.98\textwidth]{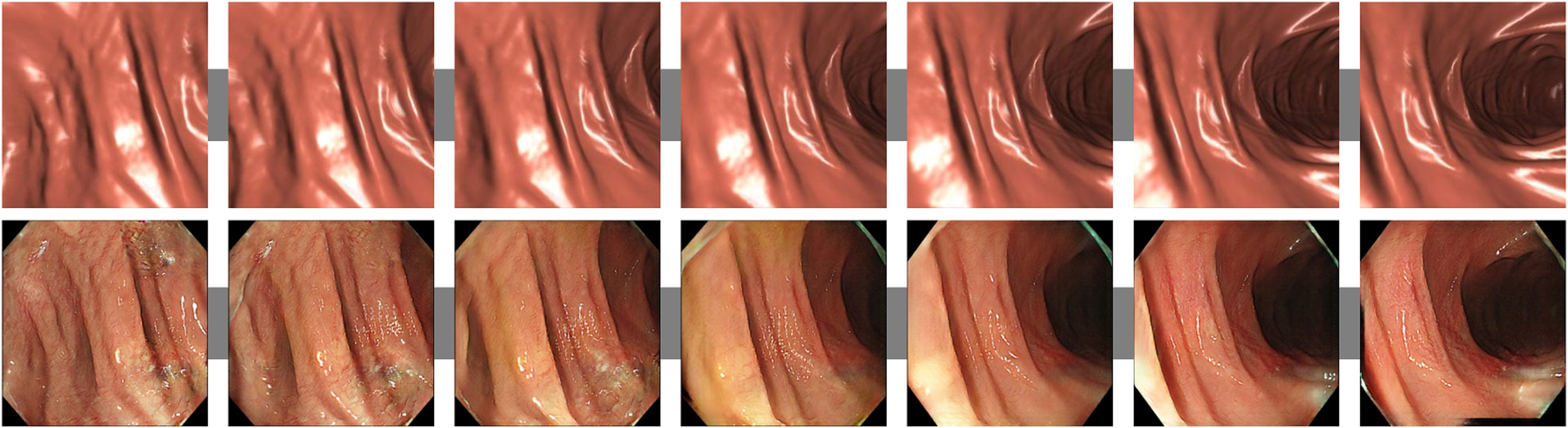} \\
(b) \\
\includegraphics[width=0.98\textwidth]{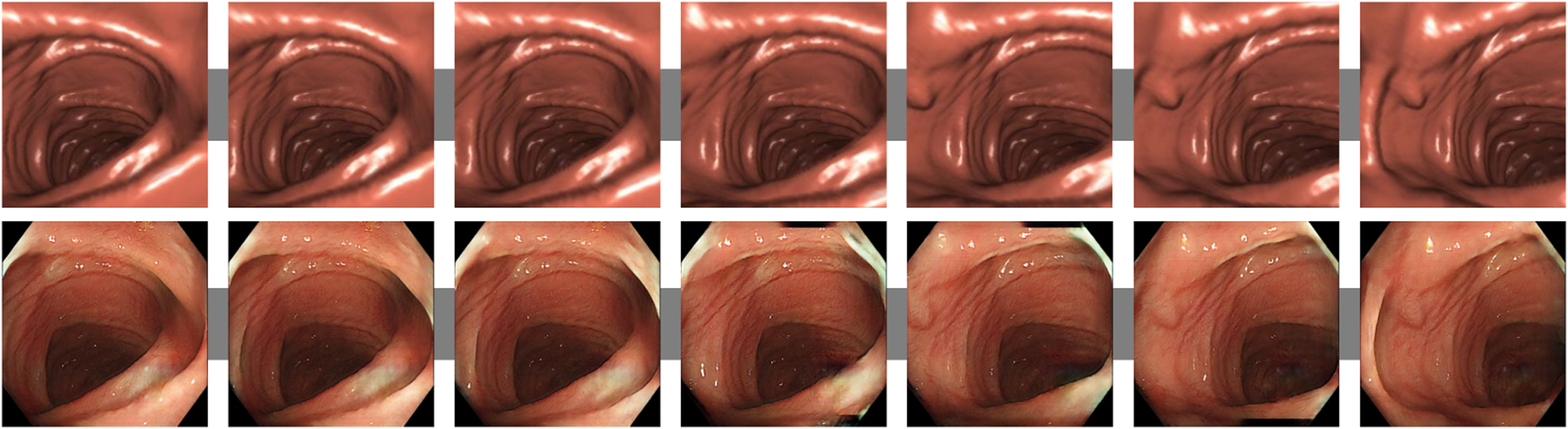} \\
(c)
\end{tabular}
\end{center}
\caption{Image-domain translation results of consecutive images in virtual colonoscopic videos. Top and bottom rows show virtual colonoscopic images and corresponding image-domain translation results using U-Net having residual units. (a), (b), and (c) show different scenes in videos.
\label{fig:result_temporal}}
\end{figure*}

By using the trained image-domain translator trained using the "Cleansing" dataset, we converted virtual colonoscopic images.
Processing times of trainings and inferences are shown in Table \ref{tab:processtime}.
Virtual colonoscopic images and corresponding image-domain translation results generated using four network structures are shown in Fig. \ref{fig:result}.
Image-domain translation results of consecutive images in virtual colonoscopic videos are shown in Fig. \ref{fig:result_temporal}.

As shown in Fig. \ref{fig:result}, virtual colonoscopic images were translated to realistic images.
Textures and light reflections on the colon surfaces in the translated images were similar to real colonoscopic images.
The arrows in Fig. \ref{fig:result} indicate unnatural parts in the translated images.
As shown in Fig. \ref{fig:result_temporal}, the translated images generated from consecutive images in videos have temporal smoothness.

The translated images generated using the deep U-Net and the U-Net having residual units were quite similar to real colonoscopic images.
Textures and light reflections (specular) on the colon surfaces were resembles to real images.
The translated images generated using the U-Net were also similar to real images.
However, an unnatural part was observed in the images (indicated by an arrow in Fig. \ref{fig:result}).
In the translated images generated using the shallow U-Net, many unnatural parts were observed (indicated by arrows in Fig. \ref{fig:result}).
Furthermore, textures and light reflections in these images were close to those of the virtual colonoscopic images.

\begin{figure*}[tb]
\centering{\includegraphics[width=0.98\textwidth]{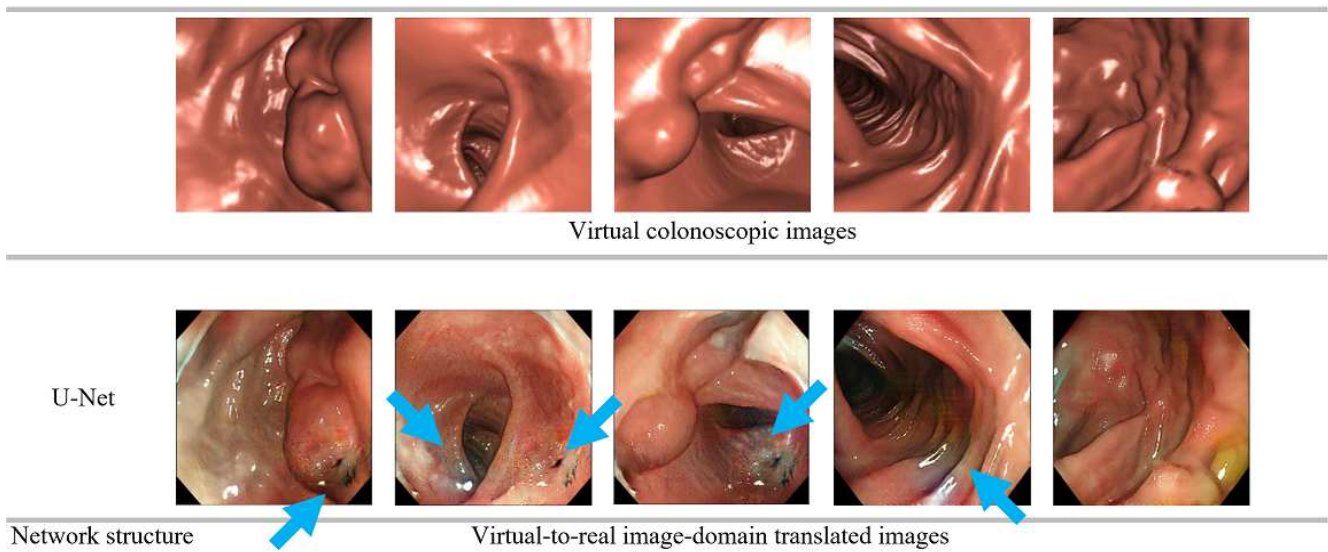}}
\caption{Top row shows virtual colonoscopic images. They were given to image-domain translator trained using "No cleansing" dataset. Bottom row shows image-domain translation results obtained using U-Net. Unnatural parts in translated images are indicated by arrows.
\label{fig:result_usenoadequate}}
\end{figure*}

\begin{figure*}[tb]
\begin{center}
\begin{tabular}{cc}
\includegraphics[width=0.4\textwidth]{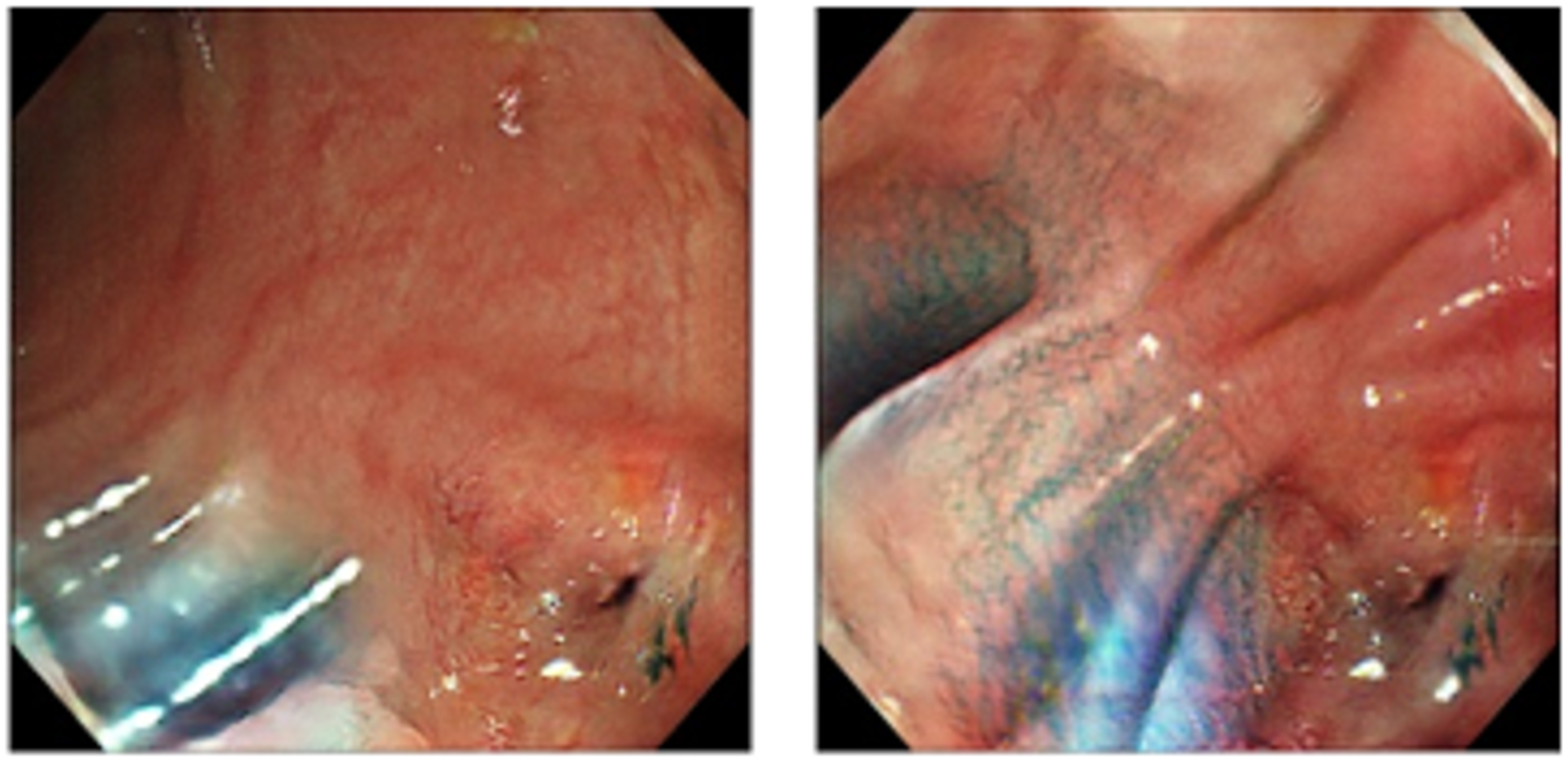} & 
\includegraphics[width=0.4\textwidth]{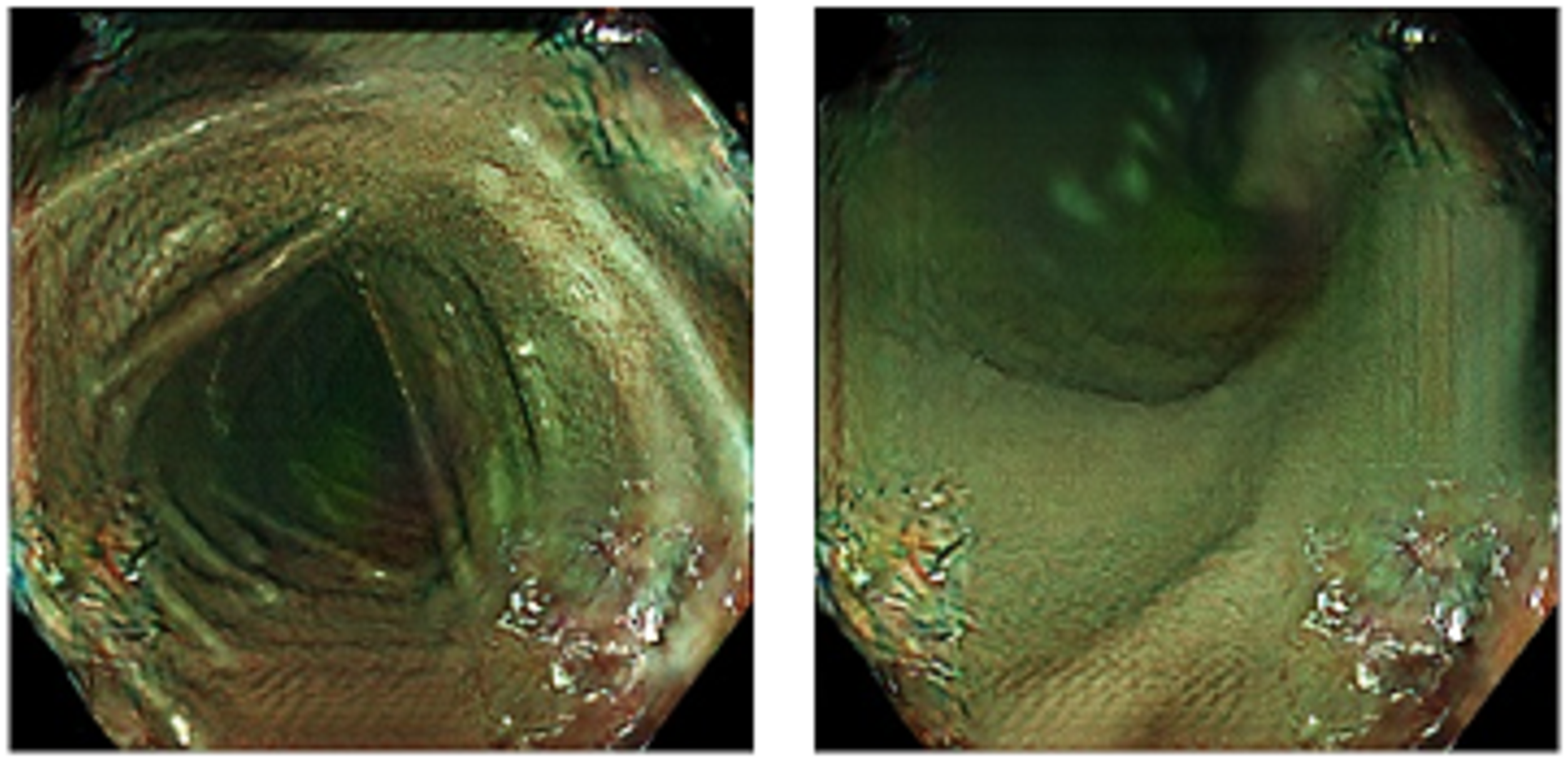} \\
(a) & (b)
\end{tabular}
\end{center}
\caption{Generated images by image-domain translator trained using "No cleansing" dataset. U-Net is used as image-domain translator. (a) contains blue haustral folds. Haustral folds were confused with blue surgical tools (sample of surgical tool is shown in Fig. \ref{fig:realimage_notusedsample}). (b) shows green textured scenes. They were affected by narrow band images (sample of image is shown in Fig. \ref{fig:realimage_notusedsample}).
\label{fig:result_usenoadequate_moresample}}
\end{figure*}

We also converted virtual colonoscopic images by using the image-domain translator trained using the "No cleansing" dataset.
The U-Net was used as the image-domain translator.
Virtual colonoscopic images and corresponding image-domain translation results are shown in Fig. \ref{fig:result_usenoadequate}.
Also, other image-domain translation results are shown in Fig. \ref{fig:result_usenoadequate_moresample}.

As shown in Figs. \ref{fig:result_usenoadequate} \ref{fig:result_usenoadequate_moresample}, many unnatural parts such as black spots, blue haustral folds, and green textured scenes were observed in the translated images.

\begin{table*}[tb]
\begin{center}
\caption{Processing times of training and inference. Trainings were performed on a NVIDIA Quadro P6000 and inferences were performed on a NVIDIA TITAN V.}
\label{tab:processtime}
\begin{tabular}{|c|c|c|}
\hline
Network structure & Training & Inference (1 image) \\ \hline\hline
Shallow U-Net & 31 hr. 30 min. & 0.0039 sec. \\ \hline
U-Net & 34 hr. 44 min. & 0.0042 sec. \\ \hline
Deep U-Net & 35 hr. 40 min. & 0.0046 sec. \\ \hline
U-Net having residual units & 46 hr. 47 min. & 0.0052 sec. \\ 
\hline
\end{tabular}
\end{center}
\end{table*}

\section{Discussion}\label{sec:discussion}

We obtained very realistic images using the deep U-Net and the U-Net having residual units as image-domain translators trained using the "Cleansing" dataset.
Textures and light reflections on the colon surfaces in these images were similar to those of real colonoscopic images.
Current endoscopic simulation systems have a problem on their visualization.
Their visualizations are not realistic because of the regardless of real textures of the organ surfaces and lack of the light reflection/absorption property information of the organ surfaces.
Current simulation systems improve reality of visualization using the texture mapping technique.
However, making good texture images for the texture mapping is difficult.
Our proposed realistic endoscope image generation method reduces the problems of the current simulation systems.
Realistic textures on the colon surfaces in the translated images greatly improved image reality.
The realistic textures were automatically generated.
Also, the appearance of light reflections in the translated images are quite similar to that in the real colonoscopic images.
Furthermore, the image-domain translators generated "black corners" that commonly appear in real endoscopic images.
%The virtual colonoscopic images have no black corners.
%The translated images have black corners similar to the real endoscopic images.
The image-domain translator learned to generate the "black corners" from real endoscopic images.
The proposed method brings great improvement of reality on visualization of endoscopic simulation systems.
Simulation systems having the realistic visualization scheme greatly contribute skill improvement of surgeons or physicians in trainings or rehearsals of endoscope manipulation.

The translated images generated using the shallow U-Net were not realistic.
Textures and light reflections on the colon surfaces in these images were not changed so much from those in virtual colonoscopic images.
That is because the layer number of the shallow U-Net was not enough to represent change of appearances between virtual and real images.
Furthermore, black noises were observed in the translated images (black noises are indicated by arrows in Fig. \ref{fig:result}).
The black noises were caused in the decoding process of images from feature values in the shallow U-Net.
The shallow U-Net cannot recover spatial information sufficiently in the decoding process.
Because of the lack of spatial information, parts of the "black corners" were generated at wrong positions in the translated images.
They were observed as the black noises.
Deeper network works well as the image-domain translator.

We performed image-domain translation of consecutive images in virtual colonoscopic videos.
As shown in Fig. \ref{fig:result_temporal}, the image-domain translator generated images having temporal smoothness.
This result is important when we utilize the proposed method as visualization of endoscopic simulation systems.
Even though the proposed method employs single 2D image-based process, the results have temporal smoothness.
The U-Net having residual units successfully recovers spatial information in each image.
It also resulted in generation of images having temporal smoothness.
Furthermore, as shown in Table \ref{tab:processtime}, inference times of the image-domain translators were rapid enough to process images in real-time.
It means the proposed method can be used for real-time visualization of simulation systems.

The data cleansing process contributed to improve the quality of image-domain translation.
As shown in Figs. \ref{fig:result_usenoadequate} and \ref{fig:result_usenoadequate_moresample}, the translated images had many unnatural parts such as black spots, blue haustral folds, and green textured areas.
They were not realistic compared to the translation results generated by using the "Cleansing" dataset.
Images removed by the data cleansing process contain black endoscopes, blue surgical tools, and narrow band images.
Use of such images in training resulted in generations of black spots, blue haustral folds, and green parts.
The data cleansing process is important to obtain better image-domain translation results.

Commonly, in the training of GANs and CycleGAN, mode collapse problem occurs sometimes.
If this problem occurs, the generator network produces limited variety of images.
In our experiments, we had no mode collapse problem.
In real colonoscopic images, there is no obvious difference of appearances among images taken by different endoscopists and images taken during different colonoscope insertions.
It means feature distribution of the images are less biased.
The images are less likely to cause mode collapse problem when they are used in training of GANs.
Also, we used the relatively large mini batch size in training.
Parameters in the networks were updated using the averaged feedback of images in a mini batch.
It reduces convergence of the network parameters under effect of specific training image.
The use of large mini batch size might be contributed to reduce mode collapse problem.

\section{Conclusions}\label{sec:conclusions}

We proposed a realistic endoscopic image generation method for endoscopic simulation systems.
Current endoscopic simulation systems generate less-realistic images.
Our proposed method generates realistic virtual images.
We generate virtual endoscopic images of a patient from a CT volume.
The virtual endoscopic images are translated to realistic images using a virtual-to-real image-domain translation technique.
The image-domain translator is implemented as a FCN.
The image-domain translator is trained to minimize the cycle consistency loss.
We employed the shallow U-Net, U-Net, deep U-Net, and U-Net having residual units as the image-domain translators.
Experiments were performed to confirm relationships between the network structures of the image-domain translator and the quality of translated images.
The deep U-Net and U-Net having residual units generated quite realistic images.
However, the shallow U-Net could not increase reality of images.
Deeper network is necessary to perform virtual-to-real image-domain translation.

Future work includes quantitative evaluation of the results, applications to other organs, utilization of other network structures as image-domain translator, and introduction of temporal information in image-domain translation.

%\ack{The authors thank *** for valuable discussions.}

\fundingandinterests{
Parts of this research were supported by the AMED Grant Numbers 18lk1010028s0401, 19lk1010036h0001, 19hs0110006h0003, the MEXT/JSPS KAKENHI Grant Numbers 26108006, 17H00867, 17K20099, and the JSPS Bilateral International Collaboration Grants.
}

\end{document}